\documentclass[letterpaper]{article} 
\usepackage{aaai24}  
\usepackage{times}  
\usepackage{helvet}  
\usepackage{courier}  
\usepackage[hyphens]{url}  
\usepackage{graphicx} 
\usepackage{natbib}  
\usepackage{caption} 
\usepackage{amsmath,amssymb} 
\usepackage{color}
\usepackage{multirow}
\usepackage{mathrsfs}
\usepackage{subfigure}
\usepackage{array}
\usepackage{threeparttable}
\usepackage{float}
\usepackage{cite}
\usepackage{algorithm}
\usepackage{algpseudocode}
\usepackage{eqparbox}
\usepackage{bm}
\usepackage{booktabs}
\usepackage{makecell}
\usepackage{diagbox}
\usepackage{picins}

\begin{document}
	\title{EEG-based Emotion Style Transfer Network for Cross-dataset Emotion Recognition}
	\author{Yijin Zhou\textsuperscript{1},
		Fu Li\textsuperscript{1},
		Yang Li$^*$\textsuperscript{1},
		Youshuo Ji\textsuperscript{1},
		Lijian Zhang\textsuperscript{2},
		Yuanfang Chen\textsuperscript{2},
		Wenming Zheng\textsuperscript{3},
		Guangming Shi\textsuperscript{1}\\
		\textsuperscript{1}{Key Laboratory of Intelligent Perception and Image Understanding of Ministry of Education, the School of Artificial Intelligence, Xidian University, Xi’an, China.\protect}\\
		\textsuperscript{2}{Beijing Institute of Mechanical Equipment, Beijing, China.\protect}\\
		\textsuperscript{3}{Key Laboratory of Child Development and Learning Science (Ministry of Education), School of Biological Sciences and Medical Engineering, Southeast University, Nanjing, Jiangsu, China.\protect }
		\it{($^*$Corresponding author: Yang Li (E-mail: liy@xidian.edu.cn).)}
	}
	\maketitle
	\begin{abstract}
		As the key to realizing aBCIs, EEG emotion recognition has been widely studied by many researchers. Previous methods have performed well for intra-subject EEG emotion recognition. However, the style mismatch between source domain (training data) and target domain (test data) EEG samples caused by huge inter-domain differences is still a critical problem for EEG emotion recognition. To solve the problem of cross-dataset EEG emotion recognition, in this paper, we propose an EEG-based Emotion Style Transfer Network (E$^2$STN) to obtain EEG representations that contain the content information of source domain and the style information of target domain, which is called stylized emotional EEG representations. The representations are helpful for cross-dataset discriminative prediction. Concretely, E$^2$STN consists of three modules, i.e., transfer module, transfer evaluation module, and discriminative prediction module. The transfer module encodes the domain-specific information of source and target domains and then re-constructs the source domain's emotional pattern and the target domain's statistical characteristics into the new stylized EEG representations. In this process, the transfer evaluation module is adopted to constrain the generated representations that can more precisely fuse two kinds of complementary information from source and target domains and avoid distorting. Finally, the generated stylized EEG representations are fed into the discriminative prediction module for final classification. Extensive experiments show that the E$^2$STN can achieve the state-of-the-art performance on cross-dataset EEG emotion recognition tasks.
	\end{abstract}
	
	\section{Introduction}
		
		As a new way of human-computer interaction in the 21st century, brain-computer interface (BCI) technology provides a new means of communication for us under the hot metaverse background~\cite{guo2022metaverse}. Emotion plays a very significant role in human-computer interaction, which has led to the great attention of affective Brain-Computer Interfaces (aBCIs) in the interdisciplinary fields~\cite{FIORINI2020105217, 7887697}. Traditional aBCIs mainly rely on two mediums, behavioral signals and physiological signals, for emotion recognition~\cite{he2020advances}. Compared with behavioral signals such facial expressions, speech, and text, it is more reliable to distinguish the spontaneous emotion state through physiological signals such as electrocardiogram (ECG), electrooculogram (EOG), electromyogram (EMG), and electroencephalogram (EEG)~\cite{zheng2017identifying, song2021variational}. Among them, EEG signals generated in the cerebral cortex are most associated with spontaneous emotional states~\cite{he2020advances, shu2017emotion}. With the development of wearable non-invasive EEG acquisition equipment in recent years, more and more researches have focused on the field of EEG-based emotion recognition.
		
		\begin{figure}[h]
			\centering
			{\includegraphics[width=1\linewidth]{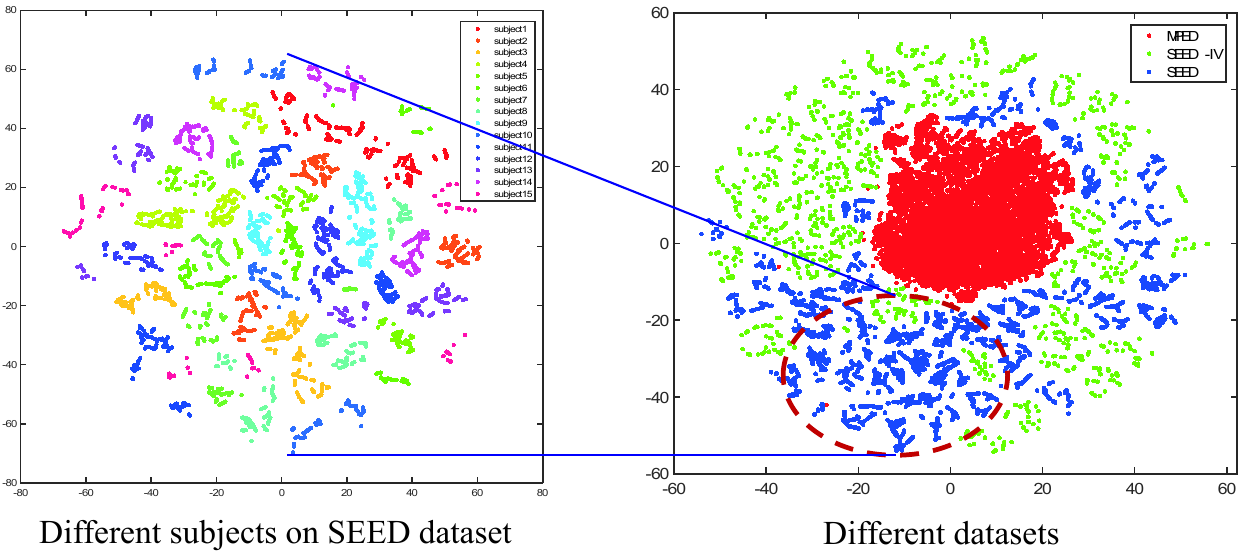}}
			\caption{\label{TSNE_domain_shift} Distribution of EEG data in different subjects and datasets.}
		\end{figure}
		
		The previous EEG emotion recognition mainly focuses on intra-subject tasks~\cite{xiao20224d}. For instance, Zheng et al. trained deep belief networks (DBNs) with differential entropy (DE) features of SEED dataset and achieved advanced results~\cite{zheng2015investigating}. Song et al. proposed the attention-long short-term memory (A-LSTM) to extract more discriminative features from multiple time-frequency domain features~\cite{song2019mped}. Peng et al. proposed a self-weighted semi-supervised classification (SWSC) model that utilizes a self-weighted variable to adaptively and quantitatively learn the importance of EEG features for cross-session emotion pattern recognition~\cite{peng2021self}. Zhong et al. proposed a regularized graph neural network (RGNN) that considers the biological topology among different brain regions to capture the local and global relationships between different EEG channels, which achieves advanced performance in cross-subject experiments on SEED and SEED-IV datasets~\cite{zhong2020eeg}. Although various EEG emotion recognition methods have been proposed in the past several years, there are still some major issues that should be well investigated to further promote EEG emotion recognition. The first issue is the protocol for EEG emotion recognition. The existing protocols for EEG emotion recognition mostly are intra-subject and cross-subject EEG classification, in which the training and test EEG data come from the same experimental environment. How the performance varies between different experimental environments should be further studied, e.g., the cross-dataset EEG emotion recognition. To clearly and intuitively show the differences in the distribution of EEG data, we use the T-SNE technology to visualize the EEG data of different subjects in different datasets, as shown in Fig~\ref{TSNE_domain_shift}. Notably, there are significant differences in the distribution of EEG data among different subjects of the same dataset, and further, which are more prominent among diverse datasets.
		
		The second issue is how to reduce the domain differences. Recently, several studies have attempted to address domain shifts in cross-subject EEG emotion recognition tasks. For example, BiDANN~\cite{li2018novel} and TANN~\cite{LI202192} proposed by Li et al. have achieved advanced performance considering the distribution differences between training data and test data in the cross-subject EEG emotion recognition task. However, the inter-domain differences in cross-dataset EEG emotion recognition go well beyond that in the cross-subject EEG emotion recognition task, as shown in Fig.~\ref{TSNE_domain_shift}. Narrowing the differences between domains will further improve cross-dataset EEG emotion recognition and the generalization to new EEG emotional data. Therefore, overcoming the significant distribution divergence of EEG signals among different datasets is particularly challenging and promising for cross-dataset EEG emotion recognition.
		
		To address the aforementioned issues, in this study, we propose an EEG-based Emotion Style Transfer Network (E$^2$STN) to obtain stylized emotional EEG representations containing emotion content of the source domain and statistical characteristics of the target domain, and meanwhile, make discriminative predictions for cross-dataset emotional EEG samples. E$^2$STN consists of three unique modules, namely the transfer module, the discriminative prediction module, and the transfer evaluation module. To realize the transfer of emotional EEG samples from the source domain to the target domain, we propose a transfer module to reorganize the emotional pattern information of the source domain and the statistical properties of the target domain into new stylized EEG representations. The first layer of the transfer module is two encoders corresponding to the source and target domains to extract their domain-specific information, namely the emotional content information of the source domain and the style statistics characteristics of the target domain. The decoder in the transfer module recombines emotional content and style characteristics to obtain stylized (transferred) emotional EEG representations. To make the stylized EEG representations contain more accurate emotional content information of the source domain and the statistical characteristic style of the target domain, we design a transfer evaluation module including content-aware loss, style-aware loss, and identity loss to constrain the process of transferring emotional EEG samples from the source domain to the target domain. The content-aware loss and style-aware loss in the transfer evaluation module ensure that the stylized representations can more precisely fuse two kinds of complementary information from source and target domains, respectively. A unique identity loss ensures that the transfer module is unbiased, i.e., the stylized representations remain unchanged when the source and target domains are the same samples. These losses are constructed from features extracted by the multi-layer convolution operations, which is inspired by~\cite{Gatys_2016_CVPR}. The multi-layer convolutions explore the latent relationship between critical frequency bands and extract spatio-temporal information from the stylized representations, respectively. Finally, to extract deep features from both original (source domain) and stylized (transferred) EEG samples for discriminative predictions, we propose a discriminative prediction module that includes a dynamic graph convolutional network and two fully connected (FC) layers. The dynamic graph convolution network extracts spatial features from two types of samples, which are further fed into FC layers to generate discriminative class labels. A cross-entropy loss with the above transfer evaluation losses jointly optimizes the entire model to realize cross-dataset EEG emotion recognition from a global perspective.
		
		Main contributions of this paper are summarized as three-fold:
		
		$ \bullet $ To our best knowledge, we propose an EEG-based Emotion Style Transfer Network (E$^2$STN) for the first time to obtain stylized emotional EEG representations for cross-dataset EEG emotion recognition. The model reorganizes the emotional content information of the source domain and the statistical characteristic style of the target domain into new stylized EEG representations, thereby further performing discriminative prediction of cross-dataset emotional EEG samples.
		
		$ \bullet $ The proposed E$^2$STN is implemented under the joint optimization of the cross-entropy loss and transfer evaluation losses. The transfer evaluation losses constrain the stylized EEG representations that can more precisely fuse two kinds of complementary information from source and target domains and avoid distorting, and meanwhile, the cross-entropy loss guides the discriminative prediction for cross-dataset EEG emotion recognition.
		
		$ \bullet $ Extensive experiments show that the proposed E$^2$STN can achieve the state-of-the-art performance on cross-dataset EEG emotion recognition tasks.
		
		The rest of this paper is organized as follows. Section II summarizes a brief overview of related studies on EEG emotion recognition. Section III specifies the proposed E$^2$STN model in detail. Section IV discusses the results of the extensive experiments conducted. Finally, this paper is concluded in Section V.
		
	\section{Related Work}
	\label{Sec:Related works}
	
	\subsection{Deep Learning in EEG Emotion Recognition}
		
		With the rapid development of deep learning, numerous studies have attempted to solve EEG emotion recognition using deep learning methods. For instance, Song et al. proposed a dynamical graph convolutional neural network (DGCNN) to recognize emotional EEG signals, which can dynamically learn the information transfer relationship between the nodes in a graph and achieve good results on SEED and DREAMER datasets~\cite{8320798}. Considering the abundant spatial information in EEG signals, Song et al. proposed to convert multi-channel EEG signals into images for EEG emotion recognition, which converts the question of EEG-based emotion recognition into image recognition. To this end, they proposed a novel EEG-to-image method and a novel graph-embedded convolutional neural network (GECNN) method. Extensive experiments on four datasets have proved the effectiveness of GECNN~\cite{9448460}. Meanwhile, Song et al. presented a novel attention-long short-term memory (A-LSTM) to extract more discriminative features from EEG sequences. A-LSTM shows advanced performance on the MPED dataset proposed in the same paper~\cite{song2019mped}. In addition, Li et al. proposed a graph-based multi-task self-supervised learning model (GMSS), which integrates multiple self-supervised tasks to learn more general representations for EEG emotion recognition. The experimental results of GMSS on SEED, SEED-IV, and MPED datasets show its advanced performance in learning more discriminative and available features for EEG emotional signals~\cite{9765326}. Zheng et al. introduced deep belief networks (DBNs) to investigate the critical frequency bands and channels of EEG signals. The experiment results show that the 4th, 6th, 9th, and 12th channels and the Gamma band are more important in emotion recognition~\cite{zheng2015investigating}. Although the above methods have achieved advanced performance in EEG emotion recognition tasks, due to the massive differences in the distribution of training and test data in practical applications, these methods are disabled to perform well.
	
	\subsection{Transfer Learning in EEG Emotion Recognition}
		
		Considering that the EEG signals in cross-subject experiments have a considerable domain shift of data distribution, numerous transfer learning methods have been widely used in cross-subject EEG emotion recognition to overcome the significant distribution divergence of EEG signals among individuals. For example, Li et al. proposed a multisource transfer learning method, which regards the existing subjects as sources and the new subject as a target to realize style transfer mapping~\cite{8675478}. Sun et al. proposed a dual-branch dynamic graph convolution-based adaptive transformer feature fusion network with adapter-finetuned transfer learning (DBGC-ATFFNet-AFTL) for EEG emotion recognition. They utilized the transfer learning method to integrate the different domain features and achieved promising performance in cross-subject emotion recognition on three datasets~\cite{9857970}. Peng et al. proposed a joint feature adaptation and graph adaptive label propagation model (JAGP) for cross-subject emotion recognition. JAGP successfully implemented the inter-domain migration by extracting and integrating the domain-invariant features~\cite{9817639}. Li et al. proposed a novel bi-hemispheres domain adversarial neural network (BiDANN) for EEG emotion recognition. Inspired by the asymmetry of the left and right hemispheres in the emotional functional regions of the brain, BiDANN maps the EEG signals of the left and right hemispheres to the discriminative feature spaces, so that the model can easily classify the data representations on SEED database.~\cite{li2018novel}. Meanwhile, Li et al. also proposed a transferable attention neural network (TANN) for EEG emotion recognition. TANN is based on the local and global attention mechanism to learn the discriminative information from emotional EEG signals. Extensive experiments on SEED, SEED-IV, and MPED datasets demonstrate the superior performance of TANN~\cite{LI202192}. However, the performance of these existing methods decreases significantly when dealing with cross-data EEG emotion recognition tasks. We will compare the representative methods among them with the proposed method in Section~\ref{Sec: Experiment}.
		
	\section{Proposed Method for Emotion Recognition}
	\label{Sec: The proposed method}
		
		\begin{figure*}[h]
			\centering {\includegraphics[width=1\linewidth]{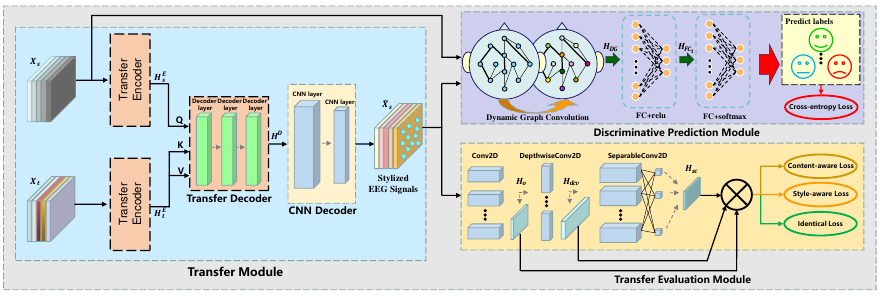}}
			\caption{\label{EESTN framework} Framework of E$^2$STN. The transfer module is applied to obtain emotional EEG representations that contain affective content information of the source domain and statistical characteristics style of the target domain. The discriminative prediction module receives stylized and source domain emotional EEG samples for cross-dataset EEG emotion recognition. Meanwhile, the transfer evaluation module progressively extracted spatio-temporal information from the stylized EEG samples to construct the multi-dimensional losses constraining the process of emotional EEG style transfer.}
		\end{figure*}
		
		To specify the proposed method clearly, we depict the E$^2$STN framework in Fig.~\ref{EESTN framework}. The network aims to re-organize the emotional content information of the source domain and the statistical characteristic style of the target domain to obtain new stylized source domain EEG samples, and finally realize the cross-dataset EEG emotion recognition task. We adopt three modules to achieve this goal, i.e., transfer module, discriminative prediction module, and transfer evaluation module. The transfer module is to obtain emotional EEG representations that contain affective content information of the source domain and statistical characteristics style of the target domain. Subsequently, the original and stylized source samples are fed into the discriminative prediction module for cross-dataset EEG emotion recognition. Meanwhile, the transfer evaluation module extracts multi-scale spatio-temporal features of the stylized EEG samples to construct the multi-dimensional losses constraining the process of emotional EEG style transfer. In the following, we introduce the details of the proposed E$^2$STN model.
		
		\subsection{Obtaining Stylized Emotional EEG Samples}
		
			To obtain emotional EEG samples that contain dynamic content information of the source domain and statistical characteristics style of the target domain, we should decompose the transfer into two steps. The first step is to construct transfer encoders corresponding to the source and target domains to capture the global dependencies in the domain-specific information of different fields (i.e., the emotional content of the source domain and the style characteristics of the target domain), respectively. Inspired by the Transformer method~\cite{NIPS2017_3f5ee243}, the encoders of E$^2$STN assign dynamic weights to different EEG channels through the multi-head self-attention layer, which selects the more important electrode dependencies in the specific domain. The dynamic dependencies with domain-specific information have stronger capabilities in representing their corresponding domain characteristics. The second key step of transfer is to fuse source domain content information and target domain style information in the decoder to obtain stylized source emotional EEG features. The decoder iteratively fuses content and style information by the multi-layer structure, which applies the target domain style to the source domain EEG features. And the residual connection method in the decoder layer ensures that the content information of the source domain will not be distorted.
			
\begin{figure}[h]
				\centering 
				\subfigure[Encoder layer]{\includegraphics[width=1\linewidth]{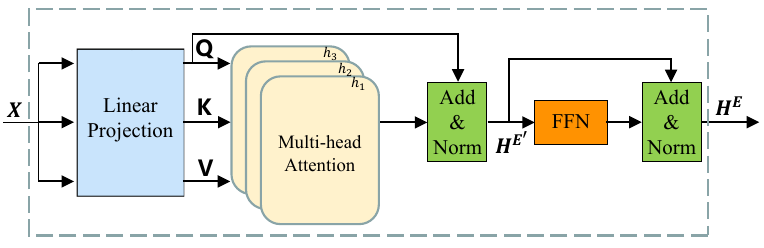}}
				\quad
				\subfigure[Decoder layer]{\includegraphics[width=1\linewidth]{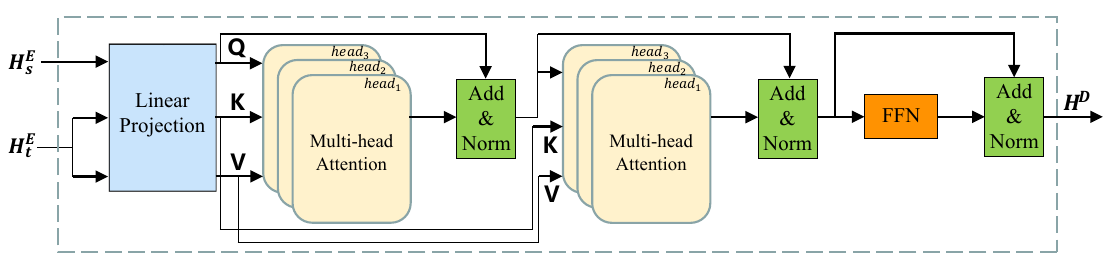}}
				\caption{\label{encoder framework} Architecture of encoder and decoder layer in transfer module.}
			\end{figure}

			Specifically, the raw EEG signals are first decomposed into five frequency bands, namely $\delta$ band (1\textendash 4 Hz), $\theta$ band (4\textendash 8 Hz), $\alpha$ band (8\textendash 14 Hz), $\beta$ band (14\textendash 30 Hz) and $\gamma$ band (30\textendash 50 Hz). The emotional EEG samples corresponding to the source and target domains, respectively, are denoted as $ {\rm \mathbf{X_s}} \in \mathbb{R} ^ {C\times B} $ and $ {\rm \mathbf{X_t}} \in \mathbb{R} ^ {C\times B} $, where $C$ is the number of EEG channels and $B$ is the number of frequency bands. Subsequently, two encoders extract the domain-specific information from their corresponding source and target domain EEG features, respectively. The encoder layer structure is depicted in Fig.~\ref{encoder framework} (a). The emotional EEG samples of the source domain $ {\rm \mathbf{X_s}} $ are first encoded into query (${\rm \mathbf{Q}}$), key (${\rm \mathbf{K}}$), and value (${\rm \mathbf{V}}$) vectors, as shown in formula~\ref{encoder 1}.
			
			\begin{equation}
				{\rm \mathbf{Q_s^E}} = {\rm \mathbf{X_s}}{\rm \mathbf{W_s^q}}, {\rm \mathbf{K_s^E}} = {\rm \mathbf{X_s}}{\rm \mathbf{W_s^k}}, {\rm \mathbf{V_s^E}} = {\rm \mathbf{X_s}}{\rm \mathbf{W_s^v}},
				\label{encoder 1}
			\end{equation}
			where $ {\rm \mathbf{W_s^q}}, {\rm \mathbf{W_s^k}}, {\rm \mathbf{W_s^v}} \in \mathbb{R} ^ {B\times m} $ are trainable linear projection matrices. To enable the encoder to pay attention to the information from different channels, $ {\rm \mathbf{Q_s^E}} $, $ {\rm \mathbf{K_s^E}} $, and $ {\rm \mathbf{V_s^E}} $ vectors are divided into several attention heads, that $ {\rm \mathbf{Q_s^E}}, {\rm \mathbf{K_s^E}}, {\rm \mathbf{V_s^E}} \in \mathbb{R} ^ {h \times C \times p} $, where $ h = \frac{m}{p} $ is the number of attention heads. Then the multi-head self-attention (MSA) can be calculated by:
			
			\begin{equation}
				{\rm \mathbf{M_s^E}} = [h_1, \cdots, h_h]{\rm \mathbf{W_s^O}} \in \mathbb{R} ^ {C\times m}.
				\label{encoder 2}
			\end{equation}
			
			To preserve domain-specific information, the MSA matrix is added with the Q vector and followed by layer normalization, which can be expressed as:
			
			\begin{equation}
				{\rm \mathbf{H_s^{E^{\prime}}}} = LN({\rm \mathbf{M_s^E}}({\rm \mathbf{Q_s^E}}, {\rm \mathbf{K_s^E}}, {\rm \mathbf{V_s^E}}) + {\rm \mathbf{Q_s^E}}),
				\label{encoder 3}
			\end{equation}
		
			\begin{equation}
				{\rm \mathbf{H_s^E}} = LN(F({\rm \mathbf{H_s^{E^{\prime}}}}) + {\rm \mathbf{H_s^{E^{\prime}}}}) \in \mathbb{R} ^ {C \times m},
				\label{encoder 4}
			\end{equation}
			
			\begin{equation}
				F(x) = max(0, x{\rm \mathbf{W_1}}+\mathbf{b_1}){\rm \mathbf{W_2}} + \mathbf{b_2},
				\label{encoder 5}
			\end{equation}
			where $ {\rm \mathbf{W_1}}, {\rm \mathbf{W_2}} $ are trainable weight matrices, and $ \mathbf{b_1}, \mathbf{b_2} $ are trainable bias matrices. Similarly, we can easily obtain the domain-specific features of the target domain $ {\rm \mathbf{H_t^E}} $.
			
			To fuse the emotional content information of the source domain and the statistical characteristic style of the target domain, we construct a three-layer Transformer decoder, which applies the style of the target domain to the emotional features of the source domain in a progressive manner. The structure of single decoder layer is shown in Fig.~\ref{encoder framework} (b). The source domain features $ {\rm \mathbf{H_s^E}} $ containing the emotional content information of the source domain are the main objects of transfer, which are used as query vectors for the first decoder layer. To make the source features more similar to the taget domain style, the target domain features $ {\rm \mathbf{H_t^E}} $ are used as key and value vectors of the first decoder layer, which calculates a similarity matrix with the query vectors to weight the emotional content features $ {\rm \mathbf{H_s^E}} $. Specifically, we obtain $ {\rm \mathbf{Q_1^D}} $, $ {\rm \mathbf{K_1^D}} $, and $ {\rm \mathbf{V_1^D}} $ through the linear projection, as shown in formula~(\ref{decoder 1}).
			
			\begin{equation}
				{\rm \mathbf{Q_1^D}} = {\rm \mathbf{H_s^E}}{\rm \mathbf{W_q^D}}, {\rm \mathbf{K_1^D}} = {\rm \mathbf{H_t^E}}{\rm \mathbf{W_k^D}}, {\rm \mathbf{V_1^D}} = {\rm \mathbf{H_t^E}}{\rm \mathbf{W_v^D}}.
				\label{decoder 1}
			\end{equation}
			Following, two MSA layers and one FFN are employed in the first decoder layer with residual connections. The output of the first decoder layer continues to be fed to the second decoder layer, and so on. Therefore, we can easily derive the output $ {\rm \mathbf{H^D}} \in \mathbb{R} ^ {C \times m} $ of the Transformer decoder through formulas~(\ref{encoder 3}),~(\ref{encoder 4}), and~(\ref{encoder 5}).
			
			To restore the dimension of the stylized features, we employ a two-layer CNN decoder to refine the output of the transformer decoder $ {\rm \mathbf{H^D}} $. Therefore, we can reshape the stylized EEG features $ {\rm \mathbf{H^D}} \in \mathbb{R} ^ {C \times m} $ as generated stylized EEG samples $ {\rm \mathbf{\hat{X}_s}} \in \mathbb{R} ^ {C \times B} $.
		
		\subsection{Obtaining Discriminative Features and Predictions}
		
			After obtaining the stylized source-domain EEG samples, we construct a dynamic graph network to extract deep features, which enables E$^2$STN to learn discriminative features from the original and stylized source samples. The source EEG samples $ {\rm \mathbf{X_s}} $ and the corresponding stylized EEG samples $ {\rm \mathbf{\hat{X}_s}} $ are jointly fed into the discriminative prediction module for obtaining discriminative features and predictions. Concretely, the data-driven graph is represented as $ \mathcal{G} = (\mathcal{V}, \mathcal{E}) $ according to the standard symbol representation in graph theory, where $ \mathcal{V} = {\{v_i\}}_{i=1}^{n} $ represents the set of $ n $ electrodes of the EEG samples, and $ (v_i,v_j) \in \mathcal{E} $ denotes the connection weight between electrodes $ v_i $ and $ v_j $ in the EEG samples. The relationship between the electrodes is characterized as an adjacency matrix $ {\rm \mathbf{G}} \in \mathbb{R}^{n \times n} $ in the E$^2$STN. The graph connection $ {\rm \mathbf{G}} $ changes adaptively with the input samples $ {\rm \mathbf{X_s}} $ and $ {\rm \mathbf{\hat{X}_s}} $. To make $ {\rm \mathbf{G}} $ contain the intra-channel spatial information and frequency band information, two trainable matrices $ {\rm \mathbf{W_s}} \in \mathbb{R} ^ {C \times C} $ and $ {\rm \mathbf{W_f}} \in \mathbb{R} ^ {B \times (C*B)} $ are multiplied left and right by input features, respectively, which can be expressed as follows:
			
			\begin{equation}
				{\rm \mathbf{G}} = ReLU[({\rm \mathbf{W_s}}{\rm \mathbf{X}} + {\rm \mathbf{B}}){\rm \mathbf{W_f}}] \in \mathbb{R} ^ {C \times (C*B)},
				\label{G_d}
			\end{equation}
			where $ {\rm \mathbf{\hat{X}_s}} = [{\rm \mathbf{X_s}}, {\rm \mathbf{\hat{X}_s}}] = \{x_1, \dots, x_t, \dots\} $, $ ReLU $ is applied to the output to guarantee non-negative elements, the bias matrix $ {\rm \mathbf{B}} \in \mathbb{R} ^ {C \times B} $ is used to increase the flexibility of graph structure representation. Then, the adjacency matrix  $ {\rm \mathbf{G}} $ is reshaped into $ B $ adjacency matrices, i.e., $ {\rm \mathbf{G}} = [{\rm \mathbf{G}_{1}^{*}}, \dots, {\rm \mathbf{G}_{B}^{*}}] \in \mathbb{R} ^ {C \times C \times B} $, to represent graphs in $ B $ frequency bands.
									
			To avoid the high computational complexity of direct graph Fourier transform based on graph filtering theories, we adopt Chebyshev polynomials to approximate the graph convolution operation~\cite{kipf2017semisupervised}. Let $ \varphi_{k}(\mathbf{G}) = \mathbf{G}^{k} $ denotes the $ k $-order polynomial of the adjacency matrix $ \textbf{G} $. Therefore, the high-level features extracted by the dynamic GCN can be expressed as follows:
			
			\begin{equation}
				{\rm \mathbf{H_{DG}}} = \mathcal{G} * \mathcal{F} = \sum_{k=0}^{K-1}\varphi_{k}({\rm \mathbf{G}}){\rm \mathbf{X}} \in  \mathbb{R} ^ {C \times F},
				\label{GCN_op}
			\end{equation}
			where $ \varphi_{k}({\rm \mathbf{G}}) $ is the $ k $-th level graph, $ F $ is the output dimension for the graph convolution operation. 
			
			Subsequently, the discriminative prediction module, as the supervision term of the E$^2$STN, applies two fully connected (FC) layers to predict the class labels. $ ReLU $ activation function is adopted to the first FC layer, and the second FC layer with $ softmax $ activation function is used to calculate the classification probability. Therefore, the output of the second FC layer $ {\rm \mathbf{H_{FC}}} = \{o_1, \dots, o_p\} \in  \mathbb{R} ^ {1 \times P} $ can be easily deduced, where $ P $ is the output dimension of the FC layer. Then, the discriminative predictions from the softmax layer for emotion recognition can be expressed as follows:
			
			\begin{equation}
				\bm{Y}(p|x_t) = exp(o_{p}) / \sum_{i=1}^{P}exp(o_{i}),
			\end{equation}
			where $ \bm{Y}(p|x_t) $ denotes the predicted probability that the input sample $ x_t $ belongs to the $ p $th class in the discriminative prediction module. Consequently, labels $ l_t $ of sample $ x_t $ are predicted as follows:
			\begin{equation}
				l_t = arg \max \limits_p \bm{Y}(p|x_t).
			\end{equation}
		
			Consequently, the cross-entropy loss function of E$^2$STN to achieve cross-dataset EEG emotion recognition can be expressed as:
			\begin{equation}
				L_{ce} = \sum_{t=1}^{M_1}\sum_{p=1}^{P}-\tau(l_g, l_t) \times log\bm{Y}(p|x_t)
				\label{coarse_loss},
			\end{equation}
			
			\begin{equation}
				\tau(x,y) =  
				\begin{cases}
					1,& if~x = y \\
					0,&otherwise
				\end{cases}.
			\end{equation}
			Here, $ l_g $ represents the ground-truth label of sample $ x_t $; $ M_1 $ is the number of training samples.
											
		\subsection{Multi-objective Joint Optimization}
			
			To optimize stylized emotional EEG samples, we specially propose a transfer evaluation module to constrain the style transfer process. We mainly consider three factors in the emotional style transfer process, and thus construct three corresponding losses, namely content-aware loss $ L_c $, style-aware loss $ L_s $, and identity loss $ L_{id} $. The first important point to consider is to preserve the emotional content information of the source domain during the transfer process. We regard the features extracted by the convolutional layer as containing the content information of the corresponding domain~\cite{Gatys_2016_CVPR}. Therefore, the content-aware loss is built from the features extracted by three unique convolutional layers in the transfer evaluation module, which can be expressed as:
			\begin{equation}
				L_c = \frac{1}{3}\sum_{i=1}^{3}\left\| f_i({\rm \mathbf{\hat{X}_s}}) - f_i({\rm \mathbf{X_s}}) \right\|_2,
				\label{loss content}
			\end{equation}
			where $ f_i(\cdot) $ denotes the convolution operation function of the $ i $-th layer in the transfer evaluation module, and $ \left\|\cdot\right\|_2 $ represents the $ \ell_2 $-norm.
			
			The second point that cannot be ignored in the transfer process is that the style characteristics of the stylized emotional EEG samples should be as similar as possible to those of the target domain. The Gram matrix of the features extracted by the convolutional layer is regarded as the statistical characteristics of the target domain~\cite{Gatys_2016_CVPR}. Therefore, We can also construct the style-aware loss for the target domain by the statistics (e.g., mean and variance) of each convolutional layer in the transfer evaluation module.
			\begin{equation}
				\begin{split}
					L_s = \frac{1}{3}\sum_{i=1}^{3}(\left\| \mu(f_i({\rm \mathbf{\hat{X}_s}})) - \mu(f_i({\rm \mathbf{X_t}})) \right\|_2 + \\
					\left\| \sigma(f_i({\rm \mathbf{\hat{X}_s}})) - \sigma(f_i({\rm \mathbf{X_t}})) \right\|_2),
				\end{split}
				\label{loss style}
			\end{equation}
			where $ \mu(\cdot) $ and $ \sigma(\cdot) $ denote the mean and variance of the features, respectively.
			
			In the last point, considering maintaining more accurate content and style information in self-style transfer, we propose an identity loss to ensure the undistorted stylized EEG samples during the progressive transfer process. Specifically, to ensure the lossless and unbiased transfer of emotional EEG samples, we input the same sample $ {\rm \mathbf{X_s}} $ ($ {\rm \mathbf{X_t}} $) into the source and target domain, and the obtained stylized emotional EEG sample $ {\rm \mathbf{\hat{X}_{ss}}} $ ($ {\rm \mathbf{\hat{X}_{tt}}} $) should be identical to the $ {\rm \mathbf{X_s}} $ ($ {\rm \mathbf{X_t}} $). Hence, the identity loss $ L_{id} $ can be defined as:
			\begin{equation}
				\begin{split}
					L_{id} = \frac{1}{3}\sum_{i=1}^{3} (\left\| f_i({\rm \mathbf{\hat{X}_{ss}}}) - f_i({\rm \mathbf{X_s}}) \right\|_2 + \left\| f_i({\rm \mathbf{\hat{X}_{tt}}}) - f_i({\rm \mathbf{X_t}}) \right\|_2),
				\end{split}
				\label{loss id}
			\end{equation}
			
			To make the features extracted by the multi-layer convolutional layers in the transfer evaluation module contain multi-dimensional and multi-scale spatio-temporal information, we employ three distinct convolution kernels to construct the convolutional network. The first 2D convolutional layer explores the latent relationship between key bands of stylized EEG features, and $ {\rm \mathbf{H_{c}}} \in \mathbb{R}^{C \times B \times F_1} $ denotes the overall output feature of the 2D convolution operation, where $ F_1 $ is the number of convolutional filters. Subsequently, the depthwise convolution is used to learn the spatial information of stylized EEG features. The depth feature $ {\rm \mathbf{H_{dc}}} \in \mathbb{R}^{1 \times B \times (F_1*D)} $ can be simply derived, where $ D $ is a depth parameter that controls the number of spatial filters in the convolution operation. In the last layer, separable convolution is the extension of depthwise convolution. On the basis of depthwise convolution, $ F_2 $ pointwise  convolutions are performed to optimally merge the spatial features. Therefore, the feature $ {\rm \mathbf{H_{dc}}} $ is further compressed in the channel dimension, $ {\rm \mathbf{H_{sc}}} \in \mathbb{R}^{1 \times B \times F_2} $. $ {\rm \mathbf{H_{c}}} $, $ {\rm \mathbf{H_{dc}}} $, and $ {\rm \mathbf{H_{sc}}} $ correspond to the features extracted by each convolutional layers in formulas~(\ref{loss content}),~(\ref{loss style}), and~(\ref{loss id}), respectively.
			
			Finally, the transfer losses in the transfer evaluation module and the cross-entropy loss in the discriminative prediction module together form a multi-objective joint optimization loss function $ L $.
						
			\begin{equation}
				L = \lambda L_c + \mu L_s + \nu L_{id} + \xi L_{ce},
			\end{equation}
			where $ \lambda, \mu, \nu, \xi $ are hyper-parameters used to control the proportion between the optimization loss functions. The E$^2$STN is optimized by iteratively minimizing $ L $, and the emphasis on transferring tasks and classification tasks is achieved by adjusting the hyperparameters. The procedure used to train the E$^2$STN is presented in Algorithm~\ref{algorithm} of Appendix A.
			
	\section{Experiments}
	\label{Sec: Experiment}
	
		\subsection{Experimental Settings}
		\label{Experiment setting}
			\subsubsection{Datasets}
			
				SEED~\cite{zheng2015investigating} is a commonly used emotion EEG dataset published by SJTU, which contains 15 subjects (7 males and 8 females) in total. Each subject in SEED was asked to attend 3 different sessions. During each session, 3 different kinds of emotional film clips (i.e., positive, neutral and negative emotions) were played in proper order. There are 5 film clips for each kind of emotion, that is, a total of 15 trials (approximately 3400 samples) are included in one session. In each session, The EEG signals and eye movements were collected with the 62-channel ESI NeuroScan System\footnote{https://compumedicsneuroscan.com/} and SMI eye-tracking glasses\footnote{https://www.smivision.com/eye-tracking/product/eye-tracking-glasses/}. The locations of the EEG electrodes are based on the international 10–20 system. The EEG data was downsampled to 200 Hz and divided into 1-s segments. The differential entropy (DE) features extracted from the downsampled EEG signals with 1-second sliding window were used as training samples. 
				
				SEED-IV~\cite{8283814} contains 15 subjects in total. Similarly, each subject was required to perform three different sessions, but 4 different kinds of emotion were triggered in each session (i.e. happy, sad, fear, and neutral). There are 6 film clips for each kind of emotion, that is, a total of 24 trials are included in one session. The dataset contains EEG and eye movement data, among which the EEG data are collected using a 62-channel ESI neuroscan system. The locations of the EEG electrodes are based on the international 10–20 system. Each session was divided into 1-s segments as a training sample. The DE features at five frequency bands were extracted from all training samples.

				The MPED~\cite{song2019mped} collects four modal physiological signals: EEG, galvanic skin response, respiration, and electrocardiogram (ECG). 30 subjects are requested to watch 28 videos, which are divided into 7 categories: joy, funny, anger, fear, disgust, sadness, and neutral emotion. Herein, only the EEG data were used for emotion recognition. To remove the noise and artifacts, a 5-order Butterworth filter was used to filter EEG data at 1-100 Hz. The processed data were decomposed into five frequency bands with a slide non-overlapping window of 1-s, and its 256-point STFT features were extracted at the same time.
			
			\subsubsection{Experiment protocol}
				
				The purpose of this paper is to study the cross-dataset EEG emotion recognition tasks. Therefore, following the principles of previous experiments, we set up groups of cross-dataset EEG emotion recognition comparison experiments. Table~\ref{The details of experimental groups settings} summarizes the setup details of the cross-dataset EEG emotion recognition experiments. The SEED dataset, which has the fewest categories of emotions of the three datasets we adopted, contains three categories: positive, negative, and neutral. consequently, to ensure the category balance of the training samples, we select the neutral, sad, and happy emotions as the emotional EEG transfer objects for SEED-IV dataset, and neutral, sad, and joy emotions for MPED dataset. The samples of all subjects from one dataset are regarded as source domain data, and the ones in another dataset are used as target domain data. In this way, we can obtain six groups for 3-category cross-dataset EEG emotion recognition experiments, and two groups for 4-category cross-dataset EEG emotion recognition experiments. The eight groups of cross-dataset EEG emotion recognition experiments are denoted as 'MPED$^3$ $\rightarrow$ SEED$^3$', 'MPED$^3$ $\rightarrow$ SEED-IV$^3$', 'SEED-IV$^3$ $\rightarrow$ MPED$^3$', 'SEED-IV$^3$ $\rightarrow$ SEED$^3$', 'SEED$^3$ $\rightarrow$ MPED$^3$', 'SEED$^3$ $\rightarrow$ SEED-IV$^3$', 'MPED$^4$ $\rightarrow$ SEED-IV$^4$', and 'SEED-IV$^4$ $\rightarrow$ MPED$^4$', respectively.
				
				\begin{table*}
					\centering
					\fontsize{8}{11}\selectfont    
					\caption{The details of experimental groups settings for cross-dataset EEG emotion recognition.}
					{
						\begin{tabular}{c|c|c|c}
							\hline
							\hline
							Number of emotional categories & The source domain & The target domain & The emotional categories \cr
							\hline
							\multirow{6}{*}{3-category}    & MPED$^3$    & SEED$^3$    & neutral, joy, sad \cr
														   \cline{2-4}
														   & MPED$^3$    & SEED-IV$^3$ & neutral, joy(happy), sad \cr
														   \cline{2-4}
														   & SEED-IV$^3$ & SEED$^3$    & neutral, happy, sad \cr
														   \cline{2-4}
														   & SEED-IV$^3$ & MPED$^3$    & neutral, happy(joy), sad \cr
														   \cline{2-4}
														   & SEED$^3$    & SEED-IV$^3$ & neutral, happy, sad \cr
														   \cline{2-4}
														   & SEED$^3$    & MPED$^3$    & neutral, joy, sad \cr
							\hline
						  	\multirow{2}{*}{4-category}    & MPED$^4$    & SEED-IV$^4$ & neutral, joy(happy), sad, fear  \cr
						  							   	   \cline{2-4}
						  								   & SEED-IV$^4$ & MPED$^4$ & neutral, happy(joy), sad, fear \cr
						  	\hline
						  	\hline
						\end{tabular}
						\label{The details of experimental groups settings}
					}
					\begin{tablenotes}
						\footnotesize
						{\item~~~~~~~~~~~~~~~~~~~~~~All labeled training data and unlabeled test data are used for cross-dataset EEG emotion recognition experiments.}
					\end{tablenotes}
				\end{table*}
			
		\subsection{Experiment Results}
		
			\subsubsection{3-category cross-dataset EEG emotion recognition}
			\label{3-category Cross-dataset EEG emotion recognition}
				To evaluate the performance of our model in cross-dataset EEG emotion recognition, we conduct extensive experiments using three datasets, namely SEED, SEED-IV, and MPED. To compare the proposed E$^2$STN with the other advanced methods of EEG emotion recognition, we also conduct the same experiments using six other methods: linear SVM~\cite{suykens1999least}, A-LSTM~\cite{song2019mped}, IAG~\cite{2020Instance}, GECNN~\cite{9448460}, BiDANN~\cite{li2018novel}, and TANN~\cite{LI202192}. We quote or reproduce their results from the literature to ensure a convincing comparison with the proposed method. The mean accuracy (ACC) and standard deviation (STD) are used as the evaluation criteria for all subjects in the test dataset. The experiment results are presented in Table~\ref{Table: cross-dataset}. The best result for each row in the Table is highlighted in boldface.
				
				\begin{table*}[h]
					\centering
					\fontsize{10}{14}\selectfont    
					\caption{3-category cross-dataset classification performance for EEG emotion recognition on SEED, SEED-IV, and MPED.}
					\resizebox{\textwidth}{16mm}
					{
						\begin{tabular}{c|c|c|c|c|c|c}
							\hline
							\hline
							{\multirow{2}{*}{Method}}  &\multicolumn{6}{c}{\textbf{ACC / STD (\%)}} \cr
							
							\cline{2-7}
							&{MPED$^3$ $\rightarrow$ SEED$^3$} &{SEED-IV$^3$ $\rightarrow$ SEED$^3$} &{MPED$^3$ $\rightarrow$ SEED-IV$^3$} &{SEED$^3$ $\rightarrow$ SEED-IV$^3$} &{SEED-IV$^3$ $\rightarrow$ MPED$^3$} &{SEED$^3$ $\rightarrow$ MPED$^3$} \cr
														
							\hline	
							{SVM~\cite{suykens1999least}} &~48.94/04.96 &~23.63/10.28 &~27.71/02.93 &~29.47/06.27 &~33.32/0.08 &~40.53/04.74 \cr
							{A-LSTM~\cite{song2019mped}} &~47.55/07.46 &~46.47/08.30 &~42.59/06.08 &~58.19/13.73 &~38.51/03.94 &~43.80/05.45 \cr
							{IAG~\cite{2020Instance}} &~60.89*/\textendash &~52.84/07.71 &~58.61/08.28 &~59.87/11.16 &~39.67/03.13 &~40.90*/\textendash \cr
							{GECNN~\cite{9448460}} &~62.90/06.58 &~58.02/07.03 &~60.88/06.96 &~57.25/07.53 &~38.82/03.52 &~43.15/03.08 \cr
							{BiDANN~\cite{li2018novel}} &~61.30/09.14 &~49.24/10.49 &~57.57/07.60 &~60.46/11.17 &~40.16/04.29 &~43.17/04.72 \cr
							{TANN~\cite{LI202192}} &~64.23/09.63 &~58.41/07.16 &~55.14/09.59 &~60.75/10.61 &~37.16/01.69 &~40.62/04.66 \cr
							{E$^2$STN} &~\textbf{73.51/07.23} &~\textbf{60.51/05.41} &~\textbf{62.32/06.60} &~\textbf{61.24/15.14} &~\textbf{40.43/04.49} &~\textbf{45.56/04.78} \cr
							\hline
							\hline
						\end{tabular}
					}
					\begin{tablenotes}
						\footnotesize
						{\item~~~ * indicates the results are obtained from the literature. The rest are obtained by our own implementation.}
					\end{tablenotes}
					\label{Table: cross-dataset}
				\end{table*}
			
				Table~\ref{Table: cross-dataset} shows that the proposed E$^2$STN model achieves the best performance in cross-dataset EEG emotion recognition experiments, thus verifying the transfer and recognition effectiveness of the proposed method. Specifically, E$^2$STN performs best on the 'MPED$^3$ $\rightarrow$ SEED$^3$' task in the the 3-category cross-dataset EEG emotion recognition experiments, with an accuracy rate of 73.51\%, which is significantly higher than that of the compared advanced algorithms. Compared with the state-of-the-art domain adaptation method TANN~\cite{LI202192}, E$^2$STN improves the accuracy by 09.28\% (73.51\% vs 64.23\%) on the 3-category classification of the 'MPED$^3$ $\rightarrow$ SEED$^3$' task. For the 'SEED-IV$^3$ $\rightarrow$ SEED$^3$' experiment, E$^2$STN achieves the highest classification accuracy with the smallest standard deviation, which proves that the proposed method has the best performance and stability. In 'MPED$^3$ $\rightarrow$ SEED-IV$^3$' and 'SEED$^3$ $\rightarrow$ SEED-IV$^3$' experiments (i.e., the 4th and 5th columns of Table~\ref{Table: cross-dataset}), the recognition performance of the proposed method has declined, which may be due to the reduction of emotional feature discrimination when the SEED-IV dataset collect finer emotion. Meanwhile, the similar classification performance of the two tasks (62.32\% vs. 61.24\%) proves the effectiveness of the proposed method in eliminating the domain shift problem. Also for the MPED dataset (which contains more emotion categories), the less discrimination between emotions leads to a further decline in model performance (the 6th and 7th columns of Table~\ref{Table: cross-dataset}). However, since E$^2$STN considers both the inter-domain differences between datasets and the dynamic connection relationship of the emotional functional regions of the brain, it still has the highest recognition performance compared with other advanced methods.
								
				To verify the confidence of our experimental results, we perform the t-test statistical analysis~\cite{hanusz2016shapiro} on each reproduced accuracy result. First, the Shapro-Wilk test (S-W test)~\cite{semenick1990tests} is performed to eliminate the accuracy data that do not follow the normal distribution hypothesis. The statistical analyses are conducted by the SPSS software (IBM SPSS Statistics\footnote{https://spss.mairuan.com}), and the significance level of paired t-test~\cite{hanusz2016shapiro} is defined as $ p < 0.05 $. Taking the GECNN method as an example, our proposed E$^2$STN shows significantly better ($ p < 0.05 $) with it in each cross-dataset task. From this statistical analysis, it can be seen that our proposed method can effectively reduce the inter-domain differences among different datasets and achieve efficient EEG emotion recognition.
				
\begin{figure}[h]
					\centering 
					\subfigure[MPED$^3$ $\rightarrow$ SEED$^3$]{\includegraphics[width=0.45\linewidth]{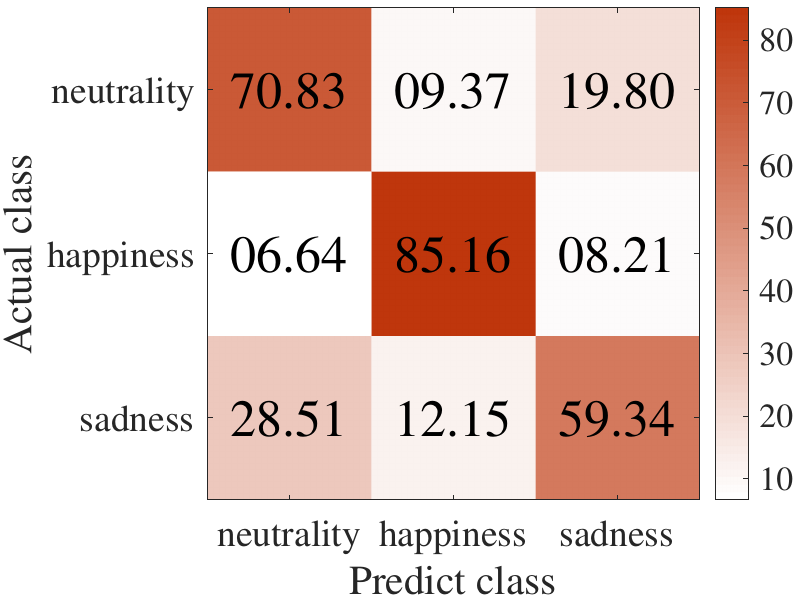}}
					\subfigure[SEED-IV$^3$ $\rightarrow$ SEED$^3$]{\includegraphics[width=0.45\linewidth]{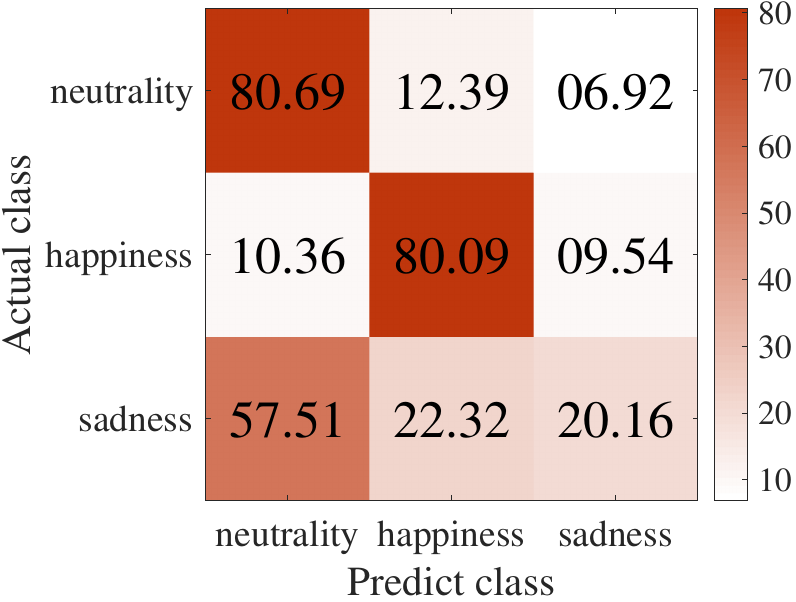}}
					\quad
					
					\centering 
					\subfigure[MPED$^3$ $\rightarrow$ SEED-IV$^3$]{\includegraphics[width=0.45\linewidth]{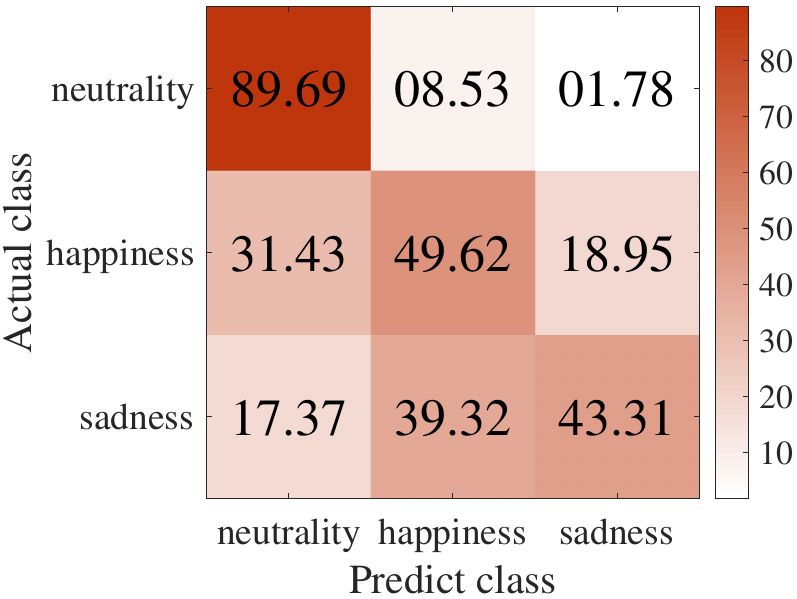}}
					\subfigure[SEED$^3$ $\rightarrow$ SEED-IV$^3$]{\includegraphics[width=0.45\linewidth]{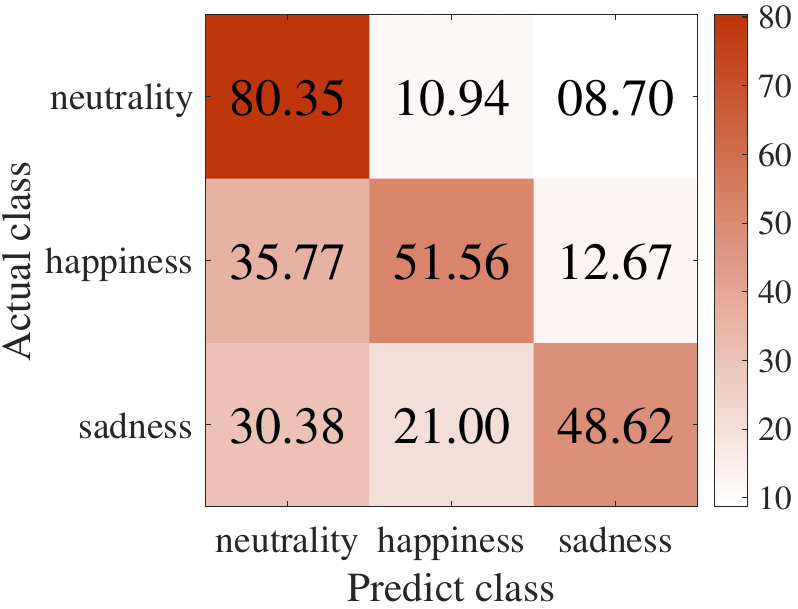}}
					\quad
					
					\centering 
					\subfigure[SEED-IV$^3$ $\rightarrow$ MPED$^3$]{\includegraphics[width=0.45\linewidth]{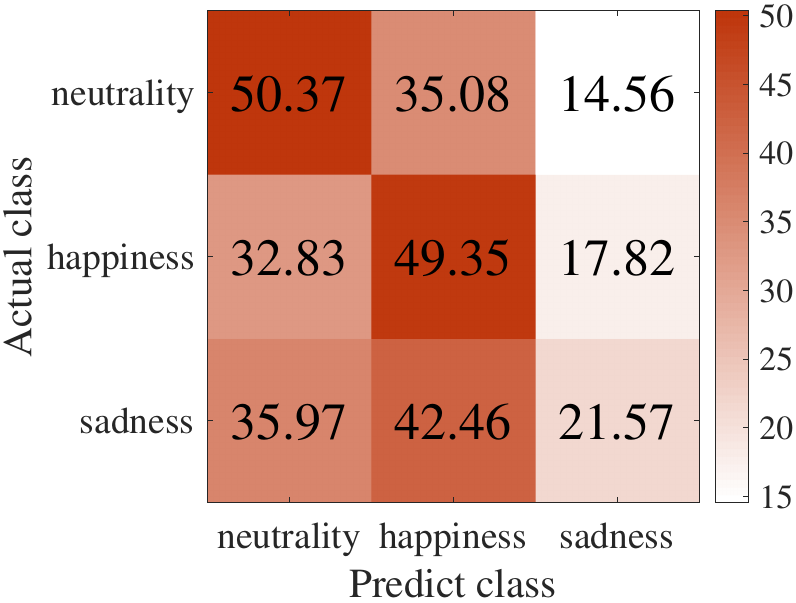}}
					\subfigure[SEED$^3$ $\rightarrow$ MPED$^3$]{\includegraphics[width=0.45\linewidth]{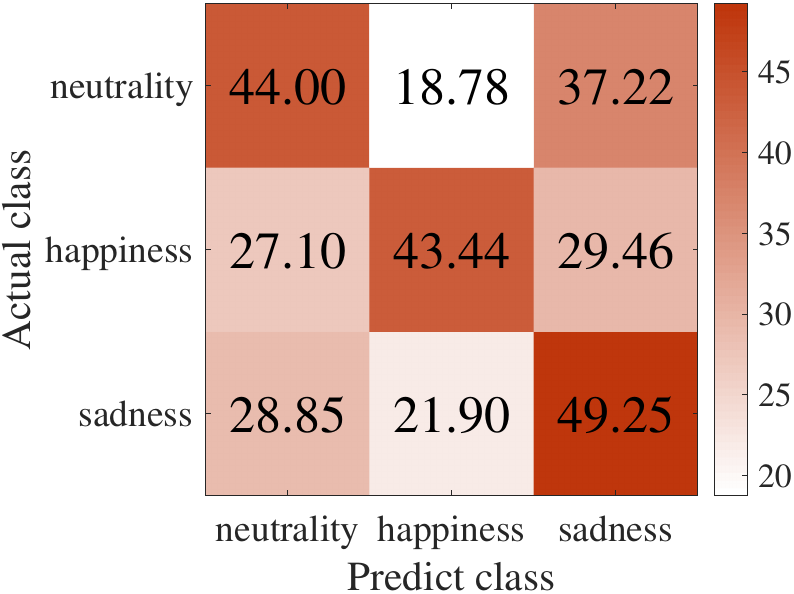}}
					\quad
					\caption*{(1) 3-category}
					
					\centering 
					\subfigure[MPED$^4$ $\rightarrow$ SEED-IV$^4$]{\includegraphics[width=0.45\linewidth]{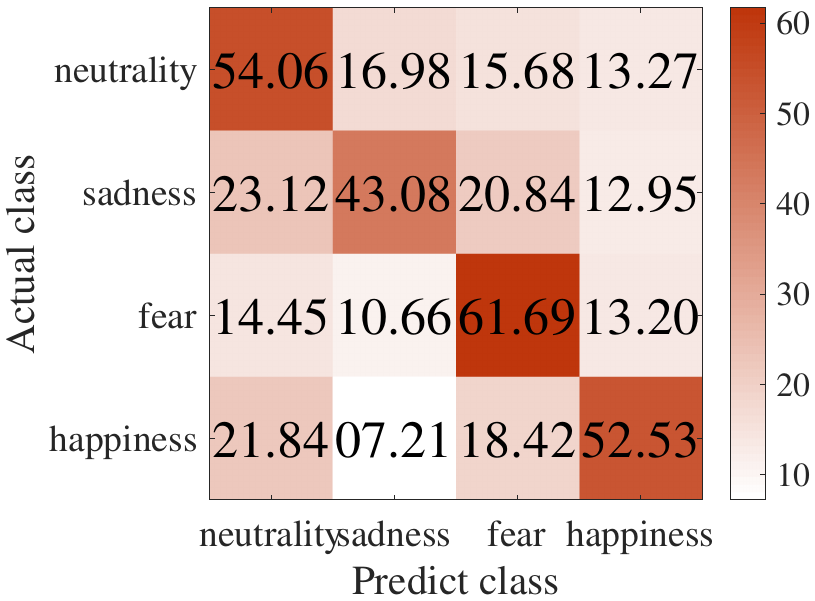}}
					\subfigure[SEED-IV$^4$ $\rightarrow$ MPED$^4$]{\includegraphics[width=0.45\linewidth]{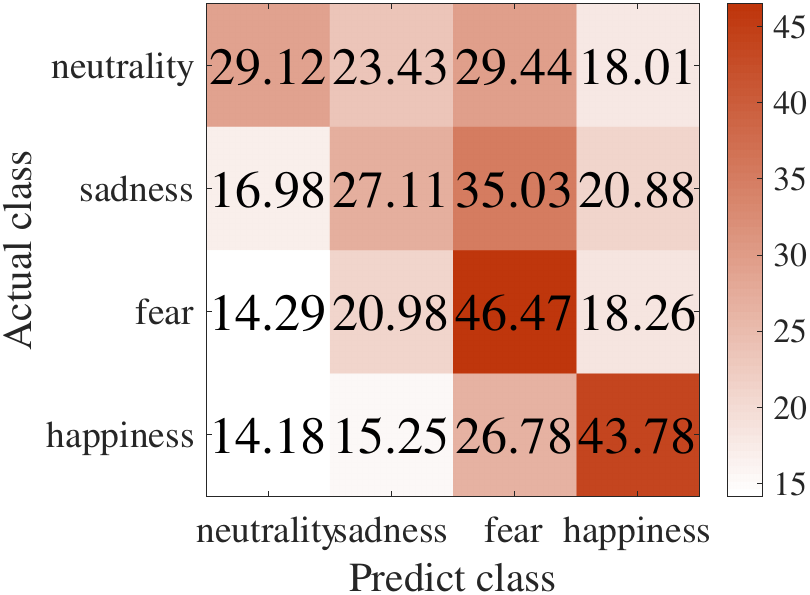}}
					\quad
					\caption*{(2) 4-category}
					
					\caption{\label{Confusion matrix}Confusion matrices of E$^2$STN results on cross-dataset experiments.}
				\end{figure}

				To explore which emotion is more easily recognized by the proposed model, we depict confusion matrices based on the results of the E$^2$STN, which are shown in Fig.~\ref{Confusion matrix} (1). From this figure, we can have the following observations. Except for the 'SEED$^3$ $\rightarrow$ MPED$^3$' experiment (Fig.~\ref{Confusion matrix} (f)), the recognition accuracy of 'happiness' emotion is higher than that of 'sadness' emotion, with an average of 24.56\% higher. This proves 'happiness' is more accessible to distinguish than 'sadness' emotions, indicating that the 'happiness' emotion is more easily induced in different datasets. Furthermore, compared with the 'happiness' emotion, the average accuracy of 'sadness' emotion is 40.38\%. The lower recognition accuracy of 'sadness' emotion is because it is easy to be mistaken for the 'neutrality' emotion, especially in Fig.~\ref{Confusion matrix}(a), (b), (d), and (f), which may be because the "sadness" emotion is weakly stimulated in these experiments. For the 'MPED$^3$ $\rightarrow$ SEED-IV$^3$' and 'SEED$^3$ $\rightarrow$ SEED-IV$^3$' experiments, we can have similar observation between Fig.~\ref{Confusion matrix}(c) and (d) with the SEED-IV dataset as the target dataset. Namely, the recognition accuracies of E$^2$STN for the three emotions have the following relationship: 'neutrality' \textgreater 'happiness' \textgreater 'sadness'. The increase in emotional categories of the SEED-IV dataset may result in more subtle emotional changes thus more difficult to recognize. For the 'SEED$^3$ $\rightarrow$ MPED$^3$' experiment in Fig.~\ref{Confusion matrix}(f), the recognition accuracy of the 'sadness' emotion is highest, which is the opposite of the other experiment results. The grander difference in the emotional categories between the SEED and MPED datasets may lead to a more significant recognition of the 'sadness' emotions.
				
			\subsubsection{4-category cross-dataset EEG emotion recognition}
			\label{4-category Cross-dataset EEG emotion recognition}
				
				To evaluate the performance of E$^2$STN under more emotional categories, we conduct additional 4-category cross-dataset EEG emotion recognition experiments on SEED-IV and MPED datasets. Similarly, to compare the E$^2$STN with other representative methods in previous studies on EEG emotion recognition, we also conduct the same experiments using the comparison methods, i.e. linear SVM~\cite{suykens1999least}, A-LSTM~\cite{song2019mped}, IAG~\cite{2020Instance}, GECNN~\cite{9448460}, BiDANN~\cite{li2018novel}, and TANN~\cite{LI202192}. We reproduce their results from the literature to ensure a convincing comparison with the proposed method. The mean accuracy (ACC) and standard deviation (STD) of all subjects in the test dataset as the evaluation criteria are shown in Table~\ref{Table: 4-categories cross-dataset}. 
				
				\begin{table}[h]
					\centering
					\fontsize{10}{14}\selectfont    
					\caption{4-category cross-dataset classification performance for EEG emotion recognition on SEED-IV and MPED.}
					\resizebox{0.48\textwidth}{15mm}
					{		
						\begin{tabular}{c|cc}
							\hline
							\hline
							\multirow{2}{*}{Method} &\multicolumn{2}{c}{\textbf{ACC / STD (\%)}} \cr
							\cline{2-3}
						 	& \textbf{MPED$^4$ $\rightarrow$ SEED-IV$^4$} &\textbf{SEED-IV$^4$ $\rightarrow$ MPED$^4$} \cr
							\cline{1-3}		
							SVM~\cite{suykens1999least} &~24.62/05.66 &~24.99/0.05 \cr
							A-LSTM~\cite{song2019mped} &~35.80/06.13 &~34.07/04.55 \cr
							IAG~\cite{2020Instance} &~49.30/05.85 &~33.92/04.94 \cr
							GECNN~\cite{9448460} &~50.86/08.30 &~33.13/02.65 \cr
							BiDANN~\cite{li2018novel} &~48.56/07.73 &~32.21/06.77 \cr
							TANN~\cite{LI202192} &~49.40/07.33 &~33.73/01.95 \cr
							{E$^2$STN} &~\textbf{53.75/06.82} &~\textbf{36.78/04.79} \cr
							\hline
							\hline
						\end{tabular}\vspace{0cm}
					}
					\label{Table: 4-categories cross-dataset}
				\end{table}
			
				Compared with the 3-category cross-dataset EEG emotion recognition, the increase in the number of emotional analogies leads to a decline in the recognition performance of E$^2$STN. For example, in the 'MPED$^4$ $\rightarrow$ SEED-IV$^4$' experiment, the average accuracy rate of E$^2$STN is 53.75\%; in the 'SEED-IV$^4$ $\rightarrow$ MPED$^4$' experiment, the average accuracy rate is 36.78\%. However, E$^2$STN still achieves the highest accuracy compared with the other advanced methods. Meanwhile, in the 'MPED$^4$ $\rightarrow$ SEED-IV$^4$' experiment, E$^2$STN improves the accuracy by 02.89\% compared to the state-of-the-art method GECNN~\cite{9448460}. Similarly, it improves by 02.71\% compared with the state-of-the-art method A-LSTM~\cite{song2019mped} in the 'SEED-IV$^4$ $\rightarrow$ MPED$^4$' task. Consistent with the 3-category cross-dataset EEG emotion recognition experiment, the accuracy of the MPED dataset as the target domain is slightly lower than that of the SEED-IV dataset. This may be because the MPED dataset collects more categories of emotions, so the feature differences between emotions are more subtle, making it difficult to transfer.
				
				For the 4-category cross-dataset EEG emotion recognition experiments, we also depict the confusion matrices in Fig.~\ref{Confusion matrix} (2).  For the 'MPED$^4$ $\rightarrow$ SEED-IV$^4$' and 'SEED-IV$^4$ $\rightarrow$ MPED$^4$' experiments, we can observe from Fig.~\ref{Confusion matrix}(g) and (h) that the 'happiness' and 'fear' emotions are easier to recognize. They are on average 14.03\% (61.69\% with 52.53\% vs 43.08\%) and 18.02\% (46.47\% with 43.78\% vs 27.11\%) higher than the 'sadness' emotion, respectively. Meanwhile, the 'sadness' emotion is easily confused with the 'fear' emotion, especially in Fig.~\ref{Confusion matrix}(h), which is consistent with the neuroscience research~\cite{KRAGEL2016444} that negative emotions (such as 'sadness' and 'fear') have quite similar Euclidean distances. In addition, the recognition accuracy of the 'neutrality' emotion in 'MPED$^4$ $\rightarrow$ SEED-IV$^4$' experiment is higher than that in 'SEED-IV$^4$ $\rightarrow$ MPED$^4$' experiment (54.06\% vs 29.12\%), which is reflected in the final recognition results (53.75\% vs 36.78\% in Table~\ref{Table: 4-categories cross-dataset}).
		
		\subsection{Discussion}
		\label{Discussion}
			\subsubsection{Effect of the transfer module}
				To verify the effect of the proposed transfer module, we modify the framework of E$^2$STN to only leave the discriminative prediction module, denoted by E$^2$STN-t. E$^2$STN-t adopts the same experimental protocols as E$^2$STN but trains only with the labeled source domain samples rather than source domain and stylized EEG samples. The experiment results are listed in Table~\ref{Table: E$^2$STN-t} and~\ref{Table: 4-categories E$^2$STN-t}. Compared with E$^2$STN-t, E$^2$STN significantly improves the performance of 3- and 4-category cross-dataset EEG emotion recognition experiments. In Table~\ref{Table: E$^2$STN-t}, E$^2$STN improves the recognition accuracy by an average of 05.69\%; in Table~\ref{Table: 4-categories E$^2$STN-t} of the 4-category cross-dataset experiments, the average increase is 04.56\%. This proves that stylized emotional EEG samples can effectively improve the performance of E$^2$STN for cross-dataset EEG emotion recognition, and further verifies the effectiveness of the proposed transfer module.
				
				\begin{table*}[h]
					\centering
					\fontsize{8}{11}\selectfont    
					\caption{Ablation experiments for 3-category cross-dataset EEG emotion recognition.}
					{		
						\begin{tabular}{c|c|c|c|c|c|c}
							\hline
							\hline
							\multirow{2}{*}{Method}  &\multicolumn{6}{c}{\textbf{ACC / STD (\%)}} \cr
							
							\cline{2-7}
							&{MPED$^3$ $\rightarrow$ SEED$^3$} &{SEED-IV$^3$ $\rightarrow$ SEED$^3$} &{MPED$^3$ $\rightarrow$ SEED-IV$^3$} &{SEED$^3$ $\rightarrow$ SEED-IV$^3$} &{SEED-IV$^3$ $\rightarrow$ MPED$^3$} &{SEED$^3$ $\rightarrow$ MPED$^3$} \cr
							
							\hline	
							{E$^2$STN} &~\textbf{73.51/07.23} &~\textbf{60.51/05.41} &~\textbf{62.32/06.60} &~\textbf{61.24/15.14} &~\textbf{40.43/04.49} &~\textbf{45.56/04.78} \cr
							{E$^2$STN-t} &~65.12/08.99 &~53.70/08.53 &~55.58/08.39 &~58.89/14.35 &~38.28/04.97 &~37.85/03.71 \cr
							\hline
							\hline
						\end{tabular}
					}
					\label{Table: E$^2$STN-t}
				\end{table*}
			
				\begin{table}[h]
					\centering
					\fontsize{8}{11}\selectfont    
					\caption{Ablation experiments for 4-category cross-dataset EEG emotion recognition.}
					{		
						\begin{tabular}{c|cc}
							\hline
							\hline
							\multirow{2}{*}{Method} &\multicolumn{2}{c}{\textbf{ACC / STD (\%)}} \cr
							\cline{2-3}
							& \textbf{MPED$^4$ $\rightarrow$ SEED-IV$^4$} &\textbf{SEED-IV$^4$ $\rightarrow$ MPED$^4$} \cr
							\cline{1-3}
							{E$^2$STN} &~\textbf{53.75/06.82} &~\textbf{36.78/04.79} \cr
							{E$^2$STN-t} &~47.39/07.13 &~34.03/05.36  \cr
							\hline
							\hline
						\end{tabular}\vspace{0cm}
					}
					\label{Table: 4-categories E$^2$STN-t}
				\end{table}
				
			\subsubsection{Exploring the importance of emotion-related brain regions}
\begin{figure}[h]
					\centering 
					\subfigure{\includegraphics[width=0.8\linewidth]{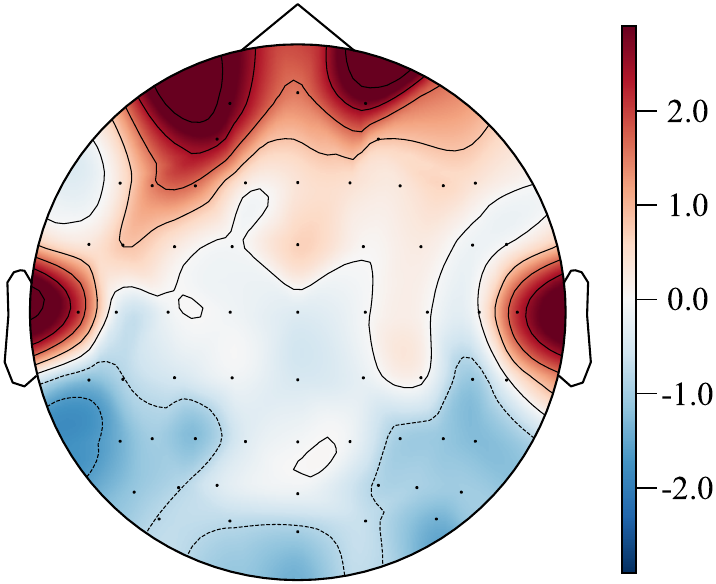}}
					\caption{\label{6}Visualization of $ {\rm \mathbf{H_{DG}}} $ distribution in the discriminative prediction module.}
				\end{figure}

				To more clearly explore the contribution of different brain functional regions for EEG emotion recognition, we depict the electrode activity maps in Fig.~\ref{6}. The contribution of each brain region is reflected in the visualization of advanced features $ {\rm \mathbf{H_{DG}}} $, which is extracted by the dynamic graph convolutional layer in discriminative prediction module. The darker red areas in the figure indicate that higher contributions from corresponding regions of the brain. It can be clearly seen that the frontal and temporal lobes of the brain are activated, which is consistent with existing neuroscience research~\cite{7946165}. This reflects that E$^2$STN extracts the most important emotion-related features in both stylized and source domain EEG samples, further demonstrating the excellent performance of the proposed method for cross-dataset EEG emotion recognition.
				
	\section{Conclusion}
	\label{Sec: Conclusion}
	
		In this study, we propose an EEG-based emotion style transfer network, called E$^2$STN, to realize effective cross-dataset EEG emotion recognition. Concretely, we design three modules to accomplish transfer, discriminative prediction, and evaluation tasks, respectively. The proposed transfer module can effectively reduce the inter-domain differences in the data distribution of different datasets and generate the stylized emotional EEG samples of the target domain. The discriminative prediction module is composed of dynamic graph convolution and fully connected layers, which is jointly trained by the source domain and stylized EEG samples to achieve accurate prediction for cross-dataset experiments. Finally, the transfer evaluation module extracts multi-scale spatio-temporal features of the stylized EEG samples to construct the multi-dimensional losses constraining the process of emotional EEG style transfer. Extensive experiments have proved the effectiveness of our proposed E$^2$STN in cross-dataset EEG emotion recognition tasks. Meanwhile, we have explored the distribution of important brain regions related to emotion, providing a basis for neurophysiology. In future research, we hope to further explore the transfer rules of emotional EEG signals, so as to further improve the performance of cross-dataset EEG emotion recognition.
		
	\newpage
	\bibliography{refbib}
	
	\newpage
	\onecolumn
	\appendix
	\renewcommand{\appendixname}{Appendix~\Alph{section}}
	\section{Appendix A: Algorithm procedure}
	
		\begin{algorithm}[h]
			\caption{Procedure of the E$^2$STN model.}
			\label{algorithm}
			\begin{algorithmic}[1]
				\Require Labeled source domain data $ {\rm \mathbf{X_s}} $, the corresponding class labels $ {\rm \mathbf{Y_s}} $; Unlabeled target domain data $ {\rm \mathbf{X_t}} $.
				\Ensure 
				\State Initialize model parameters;
				\For{each $ i \in [0,epochs] $}
				\State Calculate $ {\rm \mathbf{Q_s^E}}, {\rm \mathbf{K_s^E}}, {\rm \mathbf{V_s^E}} $ with $ {\rm \mathbf{W_s^q}}, {\rm \mathbf{W_s^k}}, {\rm \mathbf{W_s^v}} $, and $ {\rm \mathbf{X_s}} $;
				\State Calculate $ {\rm \mathbf{Q_t^E}}, {\rm \mathbf{K_t^E}}, {\rm \mathbf{V_t^E}} $ with $ {\rm \mathbf{W_t^q}}, {\rm \mathbf{W_t^k}}, {\rm \mathbf{W_t^v}} $, and $ {\rm \mathbf{X_t}} $;
				\State Conduct MSA and FFN calculations of the source and target domain, obtain the domain-specific features $ {\rm \mathbf{H_s^E}} $ and $ {\rm \mathbf{H_t^E}} $
				\Statex \ \ \ \ \ \ from the encoders;
				\For{each $ j \in [1,3] $}
				\State Calculate $ {\rm \mathbf{Q_j^D}}, {\rm \mathbf{K_j^D}}, {\rm \mathbf{V_j^D}} $ with $ {\rm \mathbf{W_q^D}}, {\rm \mathbf{W_k^D}}, {\rm \mathbf{W_v^D}} $, $ {\rm \mathbf{H_s^E}} $, and $ {\rm \mathbf{H_t^E}} $;
				\State  Conduct MSA and FFN calculations in the decoder layers, obtain the fused feature $ {\rm \mathbf{H^D}} $;
				\EndFor
				\State Refine the fused features $ {\rm \mathbf{H^D}} $ by the CNN decoder to obtain the stylized EEG features $ {\rm \mathbf{{\hat{X}_s}}} $;
				\State Make prediction by the discriminative prediction module from the source domain EEG features $ {\rm \mathbf{X_{s}}} $ and the
				\Statex \ \ \ \ \ \  corresponding stylized EEG features $ {\rm \mathbf{{\hat{X}_s}}} $, calculate the cross-entropy loss $ L_{ce} $;
				\State Extract multi-dimensional information by the transfer evaluation module, calculate transfer losses $ L_c $, $ L_s $, and $ L_{id} $.
				\State Update parameters according to gradient descent algorithm;
				\EndFor
			\end{algorithmic}
		\end{algorithm}
	
	\section{Appendix B: Implementation details}
		
		~~~Our model is trained on an NVIDIA GeForce RTX 2080Ti GPU, with CUDA 10.1 and cuDNN 7.6.2, in Tensorflow. The number of EEG channels $ C $ and frequency bands $ B $ are 62 and 5; The parameters in the transfer module are simply set to $ d_{model} = 50 $, $ h = 10 $, and the dimension of the inner-layer in FFN $ = 256 $; The parameters in the discriminative prediction module are simply set to $ K = 5 $, $ F = 128 $, $ P = 3 $, and the output dimension of the first dense layer $ = 200 $; The parameters  of in the transfer evaluation module are simply set to $ F_1 = 8 $, $ D = 8 $, $ F_2 = 8 $; The hyper-parameters $ \lambda, \mu, \nu, \xi $ are set to 2, 10, 1, 20, respectively.
	
\end{document}